\documentclass[%
reprint,
amsmath,amssymb,
prstab,
]{revtex4-1}

\usepackage{graphicx}
\usepackage{float}
\usepackage{dcolumn}
\usepackage{bm}

\begin{document}

\preprint{APS/123-QED}

\title{Octupole based current horn suppression in multi-stage bunch compression with emittance growth correction}

\author{N. Sudar, Y. Ding, Y. Nosochkov, K. Bane and Z.Zhang}
\affiliation{SLAC National Accelerator Laboratory, Menlo Park, California 94025, USA
}%

\date{\today}

\begin{abstract}
High brightness linear accelerators typically produce electron beams with peaks in the head and/or tail of the current profile.  These current horns are formed after bunch compression due to non-linear correlations in the longitudinal phase space and the higher order optics of the compressor.  It has been suggested that this higher order compression can be corrected by inserting an octupole magnet near the center of a bunch compressor.  However, this scheme provides a correlated transverse kick leading to growth of the projected emittance.  We present here a method whereby octupole magnets are inserted into two sequential bunch compressors. By tuning a $\pi$ betatron phase advance between the two octupoles, the correlated transverse kick from the first octupole can be corrected by the second, while providing a cummulative adjustment of the higher order compression.
\end{abstract}

\maketitle
\section{Introduction}
Advancements in the production of high peak brightness electron beams have revolutionized the field of ultra-fast science through the advent of the X-Ray Free Electron Laser (FEL).  The ability to reach femtosecond level X-Ray pulse durations at a growing number of FEL facilities allows for the study of molecular and atomic scale structures as well as femtosecond scale dynamical processes \cite{FEL2}.  Meeting the growing demands of the scientific community requires continued improvement in electron beam quality and repetition rate at such facilities.  

In order to achieve the electron beam peak current required to drive the FEL interaction, high brightness linear accelerators typically employ multiple 4-dipole chicane bunch compressors.  The total compression is limited by non-linear correlations in the electron beam longitudinal phase space.  These correlations stem from RF curvature\cite{RF}, longitudinal space charge (LSC)\cite{LSC}, coherent synchrotron radiation (CSR) \cite{CSR1,CSR3}, and resistive wall wakefields \cite{WAKE1,WAKE2}.  Non-linear compression from the second order energy chirp is typically adjusted with a harmonic cavity \cite{HARM}.  Additional methods can be employed to reduce these correlations, \cite{OPT1,OPT2,OPT3,OPT4,OPT5,OPT6,OPT7}.  However, if left unchecked, higher order compression can lead to the production of horns in the current profile as the head and/or tail of the electron beam are over-compressed compared to the linear compression in the core of the beam \cite{LCLS,SWISSFEL}.  

These current horns can produce significant CSR in the bunch compressor causing further distortions in the longitudinal phase space.   This correlated longitudinal energy variation can result in a correlated transverse kick leading to growth in the projected emittance, reducing FEL performance as portions of the beam will not be matched ideally to the transverse focusing lattice.  Furthermore these current horns can produce significant energy modulation from longitudinal space charge and resistive wall wakefields downstream \cite{DIMITRI}.  In the Linac Coherent Light Source-II (LCLS-II) linac the electron beam must be transported from the linac exit through a 2 km bypass line, requiring control of these collective effects, \cite{LCLSII}.  Although in a normal conducting linac with a typical beam rate of 100 Hz or lower, these current horns could be removed with collimation in a dispersive section \cite{DING}, this is not a viable option for high repetition rate superconducting linacs due to the significant increase in radiated power from beam losses.

It was shown in \cite{TESSA1,TESSA2,TESSA3}, that the growth of current horns can be suppressed by adjusting the higher order compression with an octupole magnet inserted in a bunch compressor.  This scheme relies on placing an octupole at a point of significant dispersion, providing a transverse kick correlated with the longitudinal beam coordinate.  The octupole kick is then translated into a path length difference through the remainder of the chicane, providing an adjustment of the bunch compressor's third order longitudinal dispersion, $U_{5666}$.   However, the correlated transverse kick from the octupole will generally remain imprinted on the beam after the bunch compressor causing significant growth of the projected emittance.

We present here a scheme whereby inserting octupoles in two successive bunch compressors, the projected emittance growth can be corrected.  This scheme relies on the cancellation of the octupole kick from the first bunch compressor (BC1) by the octupole in the second one (BC2).   Provided that the beam undergoes a $\pi$ betatron phase advance between octupoles, the second octupole kick can be made equal and opposite to the first one while providing an additive contribution to the total $U_{5666}$ of the system.  This allows for effective suppression of current horns without significant degradation of the transverse beam quality. 

In this paper, we present a study of this scheme using current profile shaping in the LCLS-II superconducting linac as a possible application.  Section II gives an analytical description of the longitudinal phase space transformation from an octupole inserted in a bunch compressor.  In section III we provide an analytical description of the emittance correction scheme.  Section IV shows {\footnotesize ELEGANT} \cite{ELEGANT} simulations of a potential configuration for LCLS-II.  Section V gives further simulation studies for optimization of the scheme.  Section  VI provides a discussion of alignment tolerances.

\begin{figure}
\includegraphics[scale=0.17]{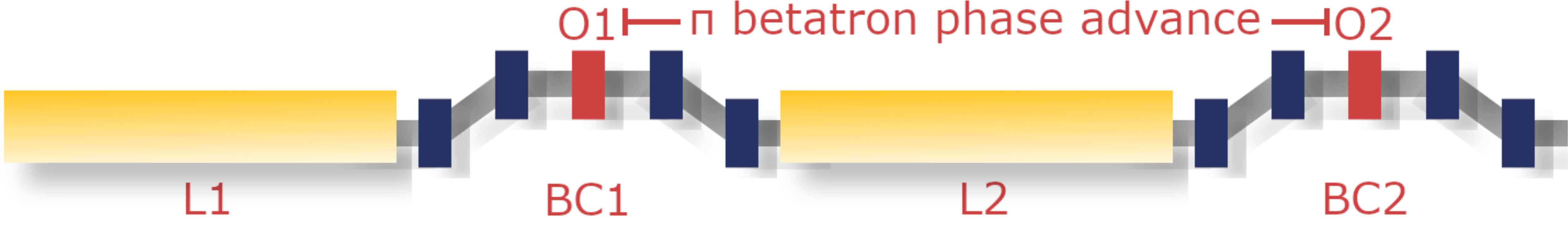}
\caption{Illustration of the proposed scheme showing the first linac section (L1), first bunch compressor (BC1) with embedded octupole (O1), second linac section (L2) and second bunch compressor (BC2) with embedded octupole (O2) }
\label{layout}
\end{figure}

\section{$U_{5666}$ from an octupole}

An electron passing through an octupole magnet with negligible vertical offset relative to the magnetic center will receive a horizontal kick depending on its horizontal offset given by \cite{CHAO}:
\begin{equation}
x' = -\frac{B'''}{6B\rho}L x^3 \equiv -\frac{1}{6}K_3 L x^3
\end{equation}
Here $L$ is the octupole length and $K_3$ is the octupole's geometric strength.  Placing an octupole at a point of high dispersion we assume an electron's transverse offset is dominated by its energy deviation from the central energy, $\delta \equiv (\gamma-\langle \gamma \rangle)/\langle \gamma \rangle$.  Placing the octupole in the center of a chicane bunch compressor, the transverse offset at the octupole entrance can be expressed by the $R_{16}$ from a simple dogleg:
\begin{equation}
x_{oct}=R_{16}\delta \approx -\theta(l_b+l_d)\delta
\end{equation}
Here $l_b$ is the chicane magnet length and $l_d$ is the drift length between the 1st and 2nd, and 3rd and 4th chicane magnets, and $\theta$ is the bending angle of the chicane dipoles.  The octupole kick then depends on the initial energy offset as:
\begin{equation}
x'_{oct} = \frac{1}{6}K_3L_0\theta^3(l_b+l_d)^3\delta^3
\end{equation}
The path length difference induced by this kick at the chicane exit is given by the $R_{52}$ from a simple dogleg.  The transformation of an electron's longitudinal position, $s$, due to the octupole kick is given by: 
\begin{equation}
\Delta s_{foct}=R_{52}x'_{oct}=-\frac{1}{6}K_3L_0\theta^4(l_b+l_d)^4\delta^3
\end{equation}
Here note we adopt the convention that the head of the beam points to more negative $s$.  The dispersive terms of the chicane including the octupole are given by:
\begin{equation}
\begin{aligned}
&R_{56}\approx -2 \theta^2(l_d+\frac{2}{3}l_b) \\
&T_{566}\approx -\frac{3}{2}R_{56} \\
&U_{5666} \approx -\frac{1}{6}K_3 L\theta^4(l_b+l_d)^4+2R_{56}
\end{aligned}
\end{equation}
We can approximate the transformation of the initial current profile first considering the evolution of the longitudinal phase space coordinates  through the chicane, $(s_i, \delta_i)\rightarrow(s_f,\delta_f)$.  For an electron beam with a non-linear correlated energy chirp described by components, $h_i$, this is given by:
\begin{equation}
\begin{aligned}
&s_i=s_0\\
&\delta_i=\delta_0+h_1 s_0+h_2{s_0}^2+h_3{s_0}^3+...\\
&s_f=s_i+R_{56}\delta_i+T_{566}{\delta_i}^2+U_{5666}{\delta_i}^3+...\\
&\delta_f=\delta_i
\end{aligned}
\end{equation}
In the limit of negligible initial energy spread the transformation of the current profile, $I_i$, can be approximated by $I_f \approx \big (\frac{\partial s_f}{\partial s_i}\big )^{-1}I_i$.  This can be expressed in terms of the linear compression factor, $C_1$ and non linear compression factors, $c_{11}$ and $c_{12}$:
\begin{equation}
I_f(s_f)\approx \frac{C_1}{\chi(s_f)}I_i[s_i(s_f)]
\end{equation}
\begin{equation}
\begin{aligned}
&C_1 \equiv \bigg (\frac{\partial s_f}{\partial s_i}\bigg|_{s_i=0}\bigg )^{-1} \\ 
&c_{11} \equiv C_1\bigg (\frac{\partial^2 s_f}{{\partial s_i}^2}\bigg|_{s_i=0}\bigg )\\
&c_{12} \equiv \frac{1}{2}C_1\bigg (\frac{\partial^3 s_f}{{\partial s_i}^3}\bigg|_{s_i=0}\bigg )\\
&\chi(s_f)=1+c_{11}s_i(s_f)+c_{12}{s_i(s_f)}^2+...
\end{aligned}
\end{equation}
Current horns will exist in the final current profile approximately where the contribution from non-linear compression, given by $\chi(s)$, goes to zero.  The adjustment of $\chi(s)$ from the octupole allows for a positive non-linear compression factor along the bunch.

Figure 2 shows an {\footnotesize ELEGANT} simulation of the nominal compression scheme for the LCLS-II beamline with a strong current horn at the beam head (I) and a hypothetical configuration with an octupole magnet inserted at the center of BC2 again simulated in {\footnotesize ELEGANT} (II).  In both cases the incoming beam is generated from {\footnotesize IMPACT} \cite{IMPACT1,IMPACT2} simulations of the LCLS-II injector.  For the single octupole case, $K_3 L_0 = -775$ $\text{m}^{-3}$ giving a total $U_{5666}=7.44$ $\text{m}^{-3}$, with remaining electron beam and chicane parameters given in Table 1.  For both cases, $\chi(s)$ is shown both as functions of the initial and final beam current. 

Here we see that in the nominal case, $\chi(s_i)$ goes to zero within the initial current profile, leading to horns in the final phase space.   The octupole effectively suppresses the current horns and can be used to produce a flat current profile with a factor of two increase in the core peak current.

Figure 2 also shows the $s$ vs $x'$ phase space at the exit of the second bunch compressor.  From this we see that the correlated octupole kick is preserved at the chicane exit.  The normalized projected emittance after BC2 is $\epsilon_{xn} = 8.84$ $\mu$m compared with $\epsilon_{xn} = 0.4$ $\mu$m for the nominal LCLS-II case. 

Assuming a gaussian energy distribution with RMS energy spread, $\sigma_\delta$, the growth of the projected emittance from the octupole kick can be approximated by:
\begin{equation}
\frac{\epsilon_x}{\epsilon_{x0}}\approx \sqrt{1+\frac{5}{12}\frac{{\beta_x}}{\epsilon_{x0}}\bigg (K_3 L (l_b+l_d)^3 \theta^3 {\sigma_{\delta}}^3\bigg )^2}
\end{equation}
Here $\epsilon_{x0}$ is the projected geometric emittance at the chicane entrance and $\beta_x$ is the electron beam beta function at the octupole.

Figure 3 shows the emittance growth versus octupole strength from {\footnotesize ELEGANT} simulations using an idealized  gaussian beam with purely linear correlated chirp.  This is done for both BC1 and BC2 using electron beam and chicane parameters given in Table 1.  Comparison with the analytical estimate from Eq. 9 shows qualitative agreement.
 
\begin{figure}[h]
\includegraphics[scale=0.32]{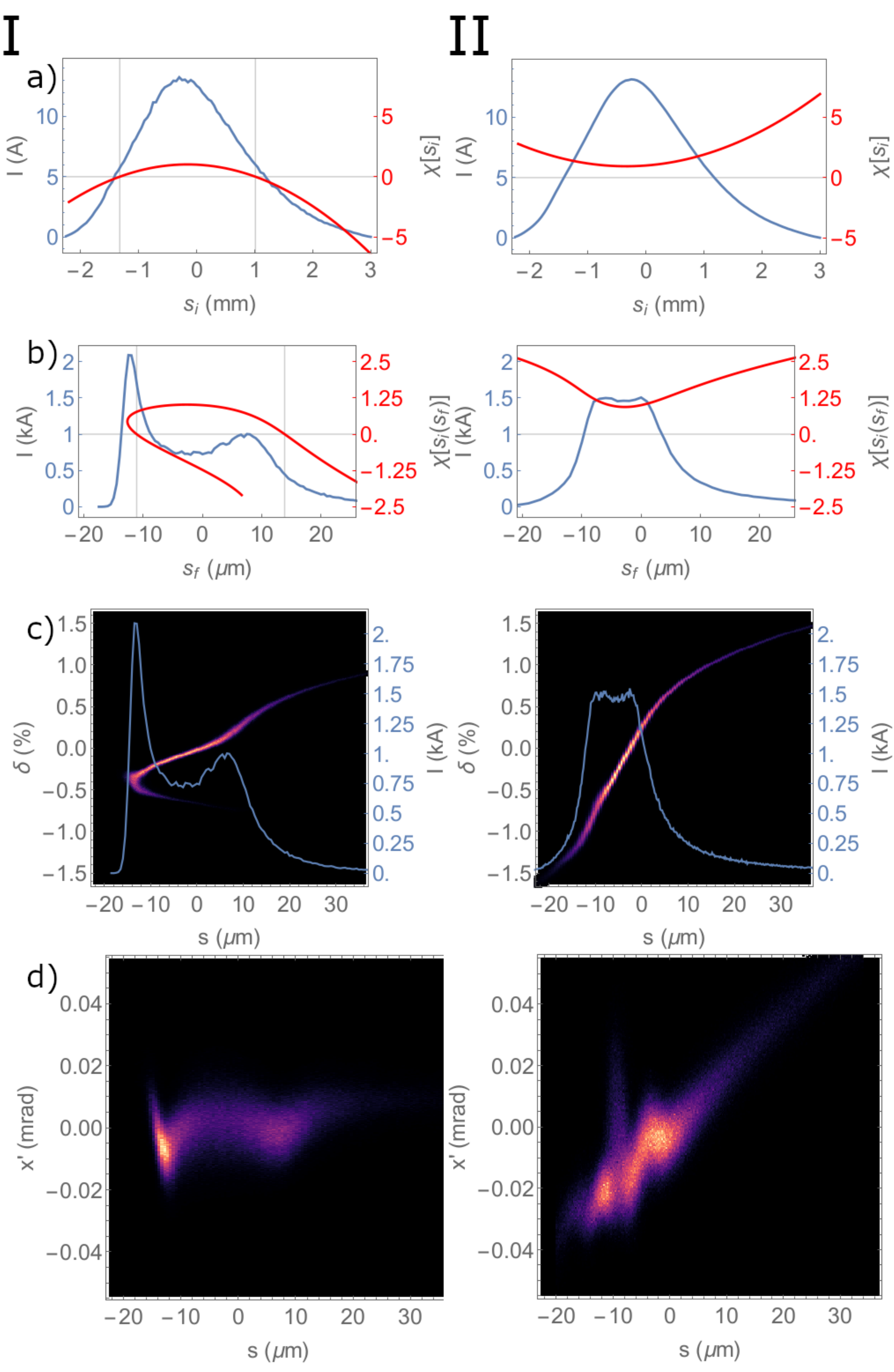}
\caption{(I) nominal LCLS-II beam (II) beam with current horns suppressed by single octupole showing: (a) Current profile (blue) and $\chi(s_i)$ (red) at linac entrance.  (b) Current profile (blue) and $\chi[s_i(s_f)]$ (red) at BC2 exit. (c) Longitudinal phase space after BC2 with current profile (blue) for reference. (d) $s$ vs $x'$ phase space at BC2 exit.  The bunch head is on the left in all plots.}
\label{layout}
\end{figure}
\begin{figure}[h]
\centering
\includegraphics[scale=0.37]{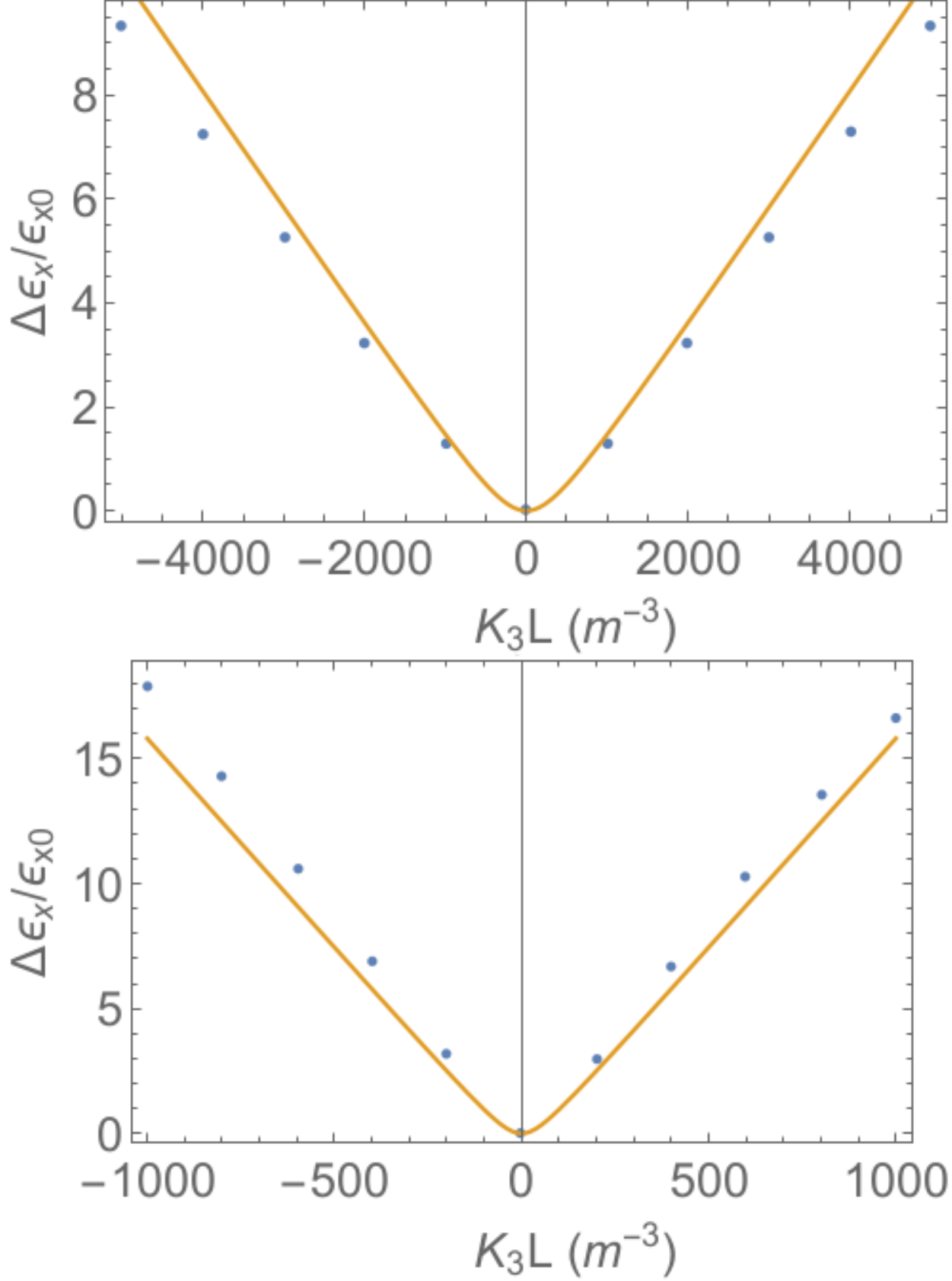}
\caption{Change in projected emittance varying octupole strength from {\footnotesize ELEGANT} simulations considering an ideal beam with a linear chirp and gaussian energy distribution (points) and the analytical expression from Eq. 9 (line) for the BC1 octupole (top) and BC2 octupole (bottom).}
\label{layout}
\end{figure}
\section{emittance correction}
The projected emittance growth problem from the single octupole scheme can be mitigated by splitting the $U_{5666}$ needed to achieve the desired longitudinal shaping between BC1 and BC2.

In this double octupole configuration, the correlated kick induced by the first octupole is transported to the second octupole.  Tuning the lattice between the two octupoles to provide a $\pi$ betatron phase advance in the bend plane of the chicanes, the $x'$ kick from the first octupole is inverted.  The second octupole strength can then be set to cancel the first octupole kick while providing additional $U_{5666}$. 

We can consider this transformation, writing the $x'$ kick at the first octupole in terms of the linear compression factor of the first chicane, $C_1$, first octupole strength and length, $K_{3}^{(1)}L_1$, chicane magnet and drift lengths and bend angle.  Writing this kick in terms of the compressed beam coordinate, $s$, after propagation through half of BC1 gives:
\begin{equation}
x'(s)=\frac{1}{6}K_{3}^{(1)} L_1\bigg(\frac{l_{d1}+l_{b1}}{l_{d1}+\frac{2}{3}l_{b1}}\bigg)^3\frac{1}{{\theta_1}^3}\bigg(\frac{C_1-1}{C_1+1} \bigg)^3 s^3
\end{equation}
Considering a more general case of $n\pi$ betatron phase advance between the two octupoles, the $x'$ kick from the first octupole at the entrance of the second octupole is given by:
\begin{equation}
\begin{aligned}
&x'(s)=(-1)^{n+1}\frac{1}{6}K_{3}^{(1)}L_1\times \\
&\bigg(\frac{l_{d1}+l_{b1}}{l_{d1}+\frac{2}{3}l_{b1}}\bigg)^3\frac{1}{{\theta_1}^3}\bigg(\frac{C_2(C_1-1)}{C_2+C_1} \bigg)^3\sqrt{\frac{\beta_1 E_1}{\beta_2 E_2}} s^3
\end{aligned}
\end{equation}
Here $C_2$ is the linear compression factor of the second chicane, $\beta_1$ and $\beta_2$ are the values of the beta function in the bend plane and $E_1$ and $E_2$ are the values of the central beam energy at the first and second octupoles respectively.

The $x'$ kick provided by the second octupole is given by:  
\begin{equation}
x'(s)=\frac{1}{6}K_{3}^{(2)}L_2\bigg(\frac{l_{d2}+l_{b2}}{l_{d2}+\frac{2}{3}l_{b2}}\bigg)^3\frac{1}{{\theta_2}^3}\bigg(\frac{C_2-C_1}{C_2+C_1} \bigg)^3 s^3
\end{equation}
The net $x'$ kick is cancelled when the second octupole kick is equal in magnitude and opposite in sign to the transported kick from the first octupole.  In reality, for LCLS-II the sign of the BC2 chicane bend angle is opposite that of BC1, requiring an $n2\pi$ betatron phase advance for the same cancellation effect.  For illustrative purposes we assume throughout that the BC1 and BC2 chicanes have the same sign of the bend angle.  In the provided example, the phase advance between the two octupoles is $3\pi$. 

The cancellation requirement gives a condition for the ratio between the two octupole strengths, $\alpha_K$:
\begin{equation}
\begin{aligned}
&\alpha_K \equiv \frac{K_{3}^{(2)}L_2}{K_{3}^{(1)}L_1}=\bigg(\frac{(l_{d1}+l_{b1})(l_{d2}+\frac{2}{3}l_{b2})}{(l_{d1}+\frac{2}{3}l_{b1})(l_{d2}+l_{b2})}\bigg)^3\bigg( \frac{\theta_2}{\theta_1}\bigg)^3\times \\
&\bigg(\frac{C_2(C_1-1)}{C_2-C_1} \bigg)^3\sqrt{\frac{\beta_1 E_1}{\beta_2 E_2}}
\end{aligned}
\end{equation}
For a given total $U_{5666}\equiv U_{tot}$ we can split the octupole strengths according to Eq. 13 and 5. 
\begin{equation}
\begin{aligned}
&K_{3}^{(1)}L_1=\frac{6U_{tot}+24{\theta_1}^2(l_{d1}+\frac{2}{3}l_{b1})+24{\theta_2}^2(l_{d2}+\frac{2}{3}l_{b2})}{{\theta_1}^4(l_{b1}+l_{d1})^4+\alpha_K {\theta_2}^4(l_{b2}+l_{d2})^4}
\end{aligned}
\end{equation}
Figure 4, shows the octupole kick from the first (a) and second (b) octupoles, the first octupole kick transported to the entrance of the second octupole (c) and the kick cancellation at the second octupole exit (d) from {\footnotesize ELEGANT} simulations of the LCLS-II linac configuration described in Section IV.  The analytical estimates from equations 10-12 are shown for comparison.  The discrepancy between the analytical expressions and simulations can be attributed to second order chromatic focusing effects in the transport between octupoles.  This is highlighted in particular by the comparison between Eq. 11 and the transported first octupole kick from simulation.  Additional discussion of the emittance correction scheme can be found in \cite{YURI}.
   
\begin{figure}[h]
\includegraphics[scale=0.32]{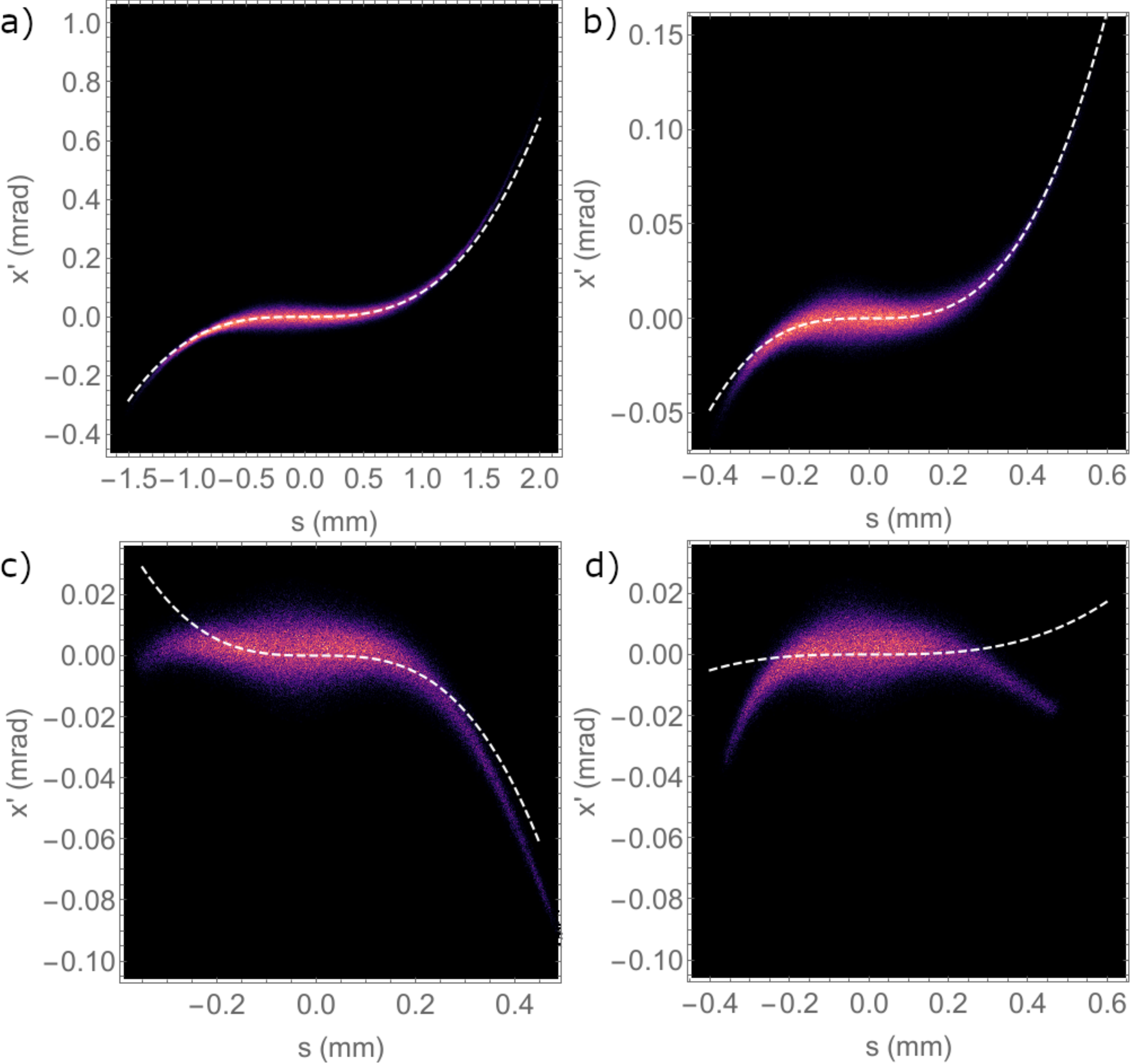}
\caption{From {\footnotesize ELEGANT} simulations: (a) Correlated kick from the first octupole ($s$ vs $x'$) with the analytical estimate from Eq. 10 (white dashed), (b) Kick from the second octupole with first octupole off  with the analytical estimate from Eq. 12 (white dashed),  (c) Kick from the first octupole transported to the second octupole entrance with the analytical estimate from Eq. 11, (d)  Combined kick from the first and second octupoles with octupole strengths chosen to minimize projected emittance of the core of the beam.  The bunch head is on the left in all plots.}
\label{layout}
\end{figure}
\section{LCLS-II double octupole configuration}
From the single octupole case shown in section II, an initial choice for the BC1 and BC2 octupole strengths is found by splitting the total $U_{5666}$ between the two chicanes according to Eq. 14.  Some adjustment of the total $U_{5666}$ must be made to obtain an identical current profile due to the change in higher order compression at BC1.  Figure 5, shows the longitudinal phase space at the exit of BC2 and undulator entrance with the double octupole configuration.  The $s$ vs $x'$ phase space at the BC2 exit shows the correction of the octupole kick.  Parameters are given in Table 1.   

\begin{table}[h]
\caption{\label{parameters} Parameters for LCLS-II double octupole scheme}
\begin{ruledtabular}
\begin{tabular}{lcdr}
\textrm{Parameter}&
\textrm{Value}\\
\colrule
BC1 e-beam energy (MeV) & 250 \\
e-beam charge (pC) & 100 \\
e-beam chirp @ BC1 entrance ($\frac{1}{m}$) & 12.97\\
emittance @ BC1 entrance $\epsilon_{x,y}$ ($\mu m$)&0.3, 0.3\\
BC1 $R_{56}$ (mm), bend angle (mrad) & -47.45, 95.66\\
BC1 compression factors $C_1,c_{11},c_{12}$ &2.6, -27.1, 2.22e4\\
BC1 octupole strength $K_{3}^{(1)}L_1$ ($m^{-3}$) & -4652.69\\
BC1 $U_{5666}$ (m) & 3.254\\
Beta function @ BC1 octupole $\beta_x$ (m) & 11.05\\
BC2 e-beam energy (MeV)& 1500 \\
e-beam chirp @ BC2 entrance ($\frac{1}{m}$) & 8.34\\
BC2 $R_{56}$ (mm), bend angle (mrad) & -44.93, 46.81\\
BC2 total compression factors $C_2,c_{21},c_{22}$ &100, 333.8, 6.2e5\\
BC2 octupole strength $K_{3}^{(2)}L_2$ ($m^{-3}$) & -511.11\\
BC2 $U_{5666}$ (m) & 4.811\\
Beta function @ BC2 octupole $\beta_x$ (m) & 55.1\\
core emittance @ BC2 exit $\epsilon_{xn,yn}$ ($\mu m$)&0.43, 0.41 \\
\end{tabular}
\end{ruledtabular}
\end{table}

\begin{figure}[h]
\centering
\includegraphics[scale=0.5]{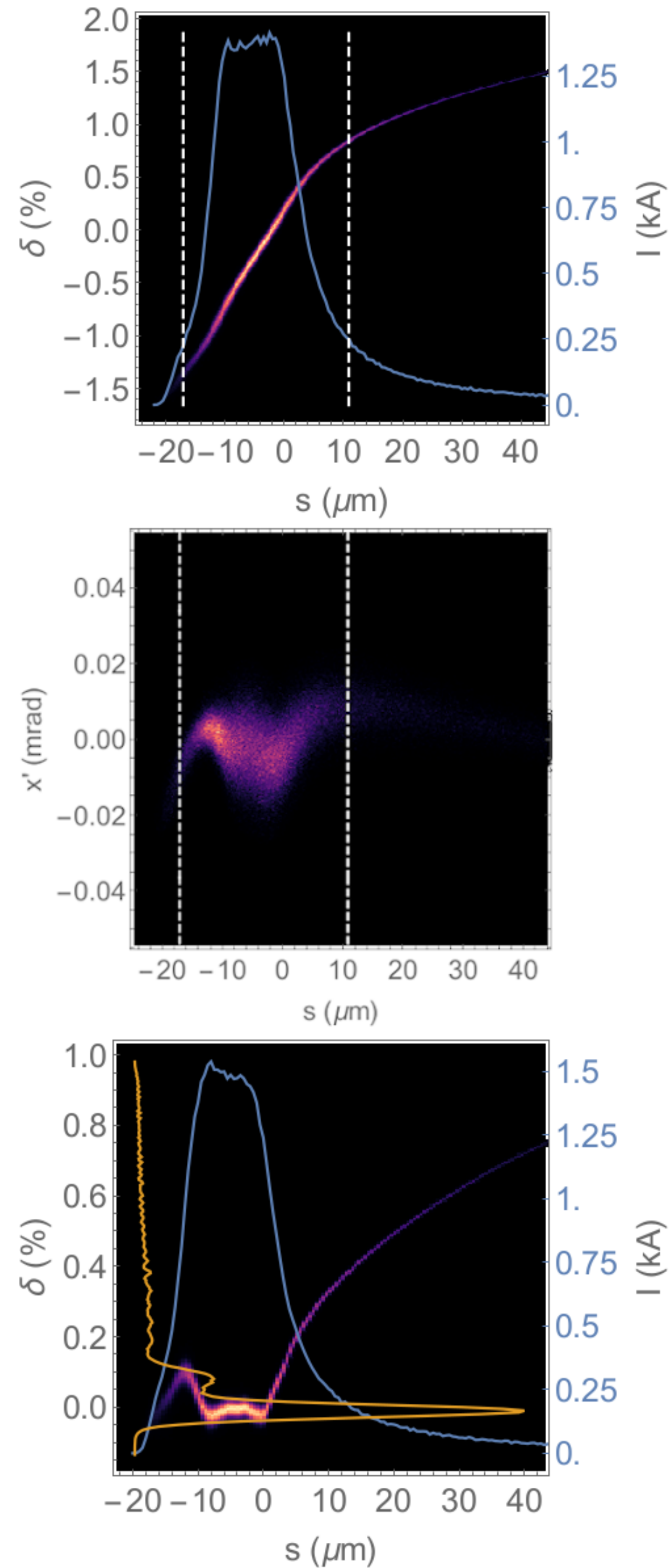}
\caption{(Top) The longitudinal phase space at the exit of BC2, with current profile (blue), the white dashed lines define the core of the beam. (Middle) $s$ vs $x'$ at the exit of BC2. (Bottom) Longitudinal phase space at the undulator entrance with current profile (blue) and energy lineout (yellow).  The bunch head is on the left in all plots.}
\label{layout}
\end{figure}

We define the core of the beam as the region within $>10\%$ of the peak current, with this region shown by the white dashed lines in Figure 5.  The ratio between octupole strengths for emittance growth cancellation can be adjusted in simulations to minimize the projected emittance over the core of the beam.   

\begin{figure}[h]
\includegraphics[scale=0.5]{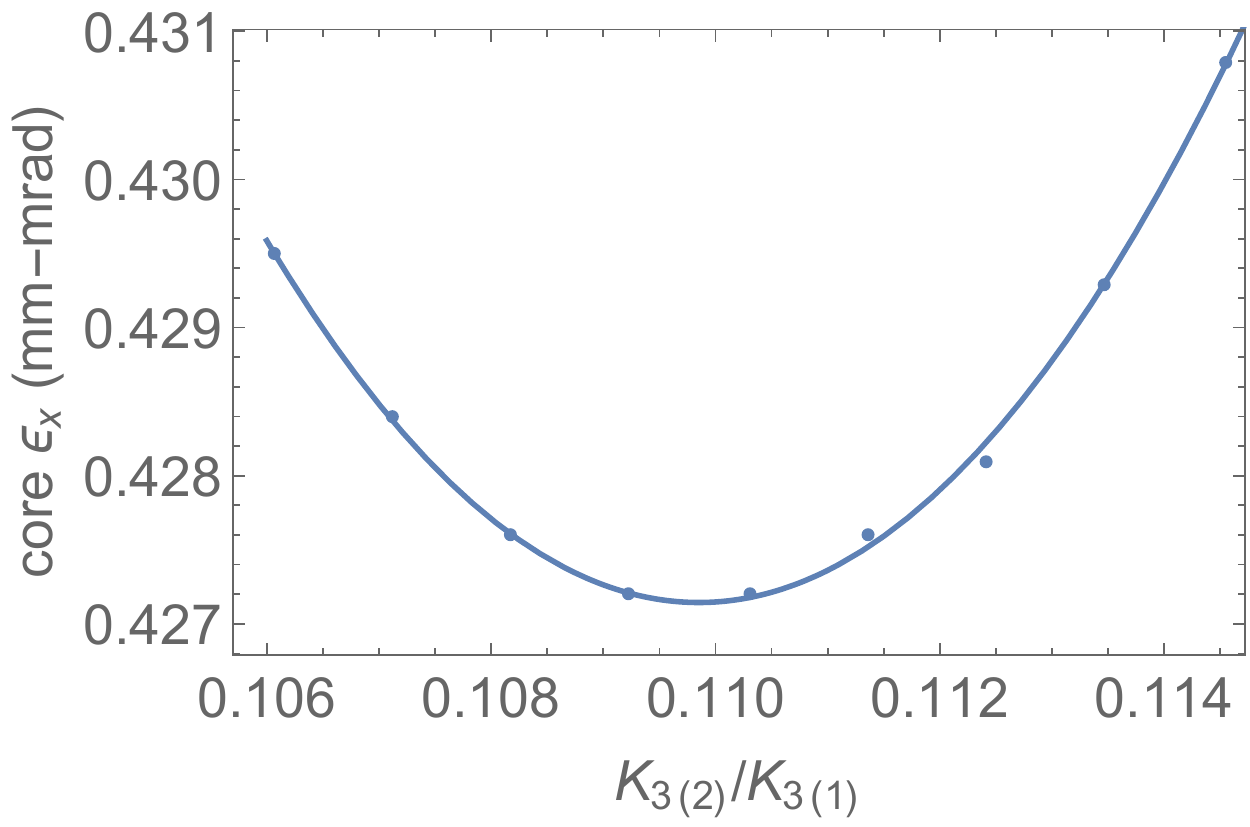}
\caption{The normalized core projected emittance varying the octupole strength ratios maintaining the total $U_{5666}$ from {\footnotesize ELEGANT} simulations (points) with quadratic fit (line)}
\label{layout}
\end{figure}

Figure 6, shows the optimal octupole strength ratio from {\footnotesize ELEGANT} simulations maintaining the total $U_{5666}$.  This gives an optimal ratio $\alpha_K = 0.1099$ showing good agreement with the analytical estimate from Eq. 13, which gives $\alpha_K = 0.098$.  After BC2 the normalized core projected emittance is $\epsilon_{xn} = 0.43$ $\mu$m compared with $\epsilon_{xn} = 0.31$ $\mu$m at the BC1 entrance and $\epsilon_{xn} = 3.1$ $\mu$m from the single octupole case.  The remnant increase in projected emittance can be again attributed to second order focusing between octupoles.  Varying $\alpha_K$ by $\pm$5\% shows negligible increase in the projected emittance.

\section{Linac configuration considerations}

In choosing a linac configuration for the double octupole scheme it is advantageous to minimize the emittance growth after the first bunch compressor.  This can reduce the remnant emittance growth.  In the case of the LCLS-II beamline, this is also necessary to avoid losses in a collimator downstream of BC1. 

We can gain some insight towards an optimal linac configuration using Eq. 9 to estimate the BC1 emittance growth for varying linac and chicane parameters that give approximately the same final current profile.  This is done by constraining the total linear compression, BC1 and BC2 non-linear compression factors (Eq. 7), the electron beam energy at BC2, and the emittance cancellation condition (Eq. 13).  With current horn suppression from the double octupole scheme, the energy modulation from linac wakefields dominates over LSC and CSR.  For changes in the BC1 compression, the linac wakefield between BC1 and BC2 can be approximately scaled by the BC1 linear compression factor. 

Figure 7 shows the estimated projected emittance growth after BC1 varying the BC1 $R_{56}$, electron beam linear chirp at the entrance of BC1, and the electron beam energy at BC1.  Here we only consider linac configurations within the specifications of the LCLS-II linac.  From this we see a general trend that the BC1 emittance growth decreases with increasing chirp and decreasing $R_{56}$.  Furthermore, for the same chirp and $R_{56}$ the emittance growth decreases with increasing electron beam energy at BC1.  The parameters for the LCLS-II example case are chosen to optimize the final phase space at the undulator entrance.  Notably, Figure 7 illustrates that the desired longitudinal shaping can be achieved over a wide range of linac parameters.  A discussion of other possible LCLS-II linac configurations is given in \cite{NICK}.   
  
\begin{figure}[h]
\includegraphics[scale=0.4]{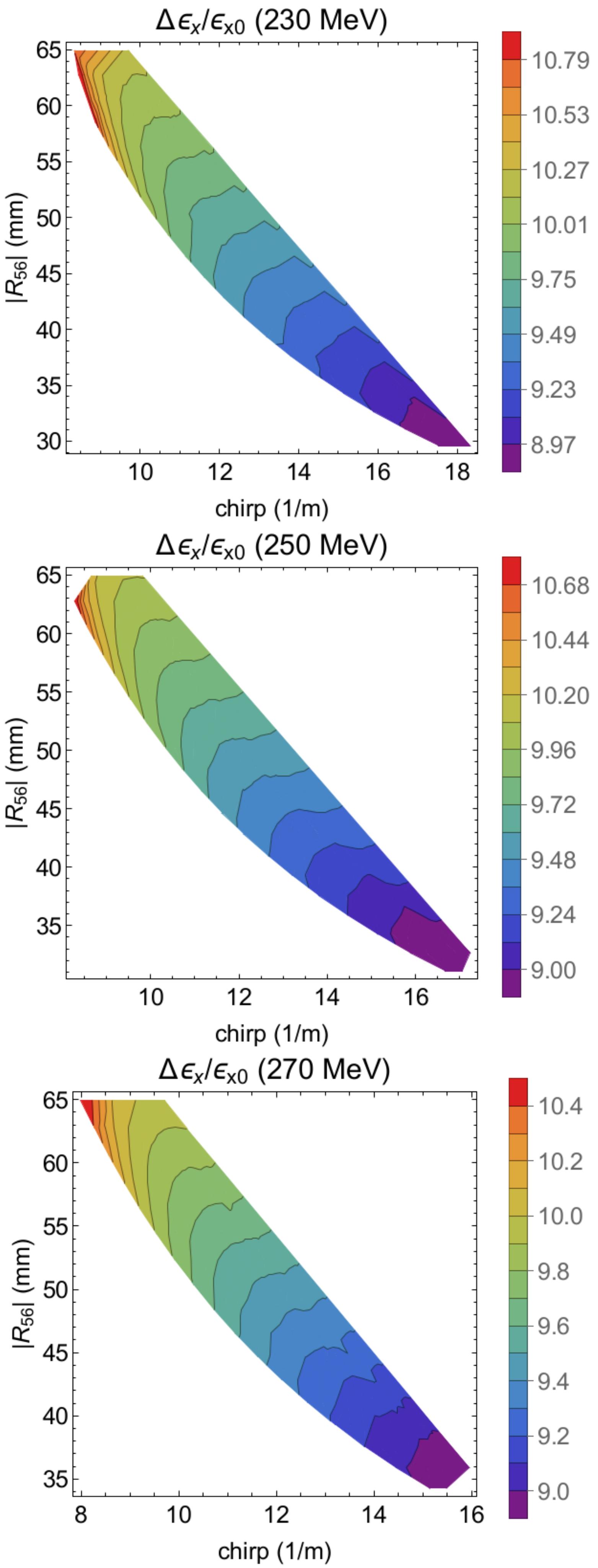}
\caption{The estimated emittance growth after BC1, varying BC1 $R_{56}$ and linear chirp at the BC1 entrance for BC1 energy 230 MeV (top), 250 MeV (middle) and 270 MeV (bottom).}
\label{layout}
\end{figure}

\section{Alignment tolerance}
Misalignment of the octupole will result in a normal and skew sextupole like kicks.  To see this we can write the octupole field shifted by $\Delta x$ and $\Delta y$:
\begin{equation}
\begin{aligned}
&B_y = \frac{B'''}{6}[(x-\Delta x)^3-3(x-\Delta x)(y-\Delta y)^2]\\
&B_x = \frac{B'''}{6}[3(x-\Delta x)^2 (y-\Delta y) -(y-\Delta y)^3]
\end{aligned}
\end{equation}
Assuming this misalignment is small compared to the dispersion and the electron beam's vertical offset relative to the magnetic center of the octupole is dominated by the vertical shift, we can write the field keeping the lowest order terms in $\Delta x$ and $\Delta y$:
\begin{equation}
\begin{aligned}
&B_y = \frac{B'''}{6}(x^3-3x^2\Delta x)\\
&B_x = -\frac{B'''}{2}(x^2 \Delta y)
\end{aligned}
\end{equation}
For the octupole inserted in the chicane, the horizontal offset at the octupole is dominated by dispersion.  The horizontal and vertical kick induced by the shifted octupole is then given by:
\begin{equation}
\begin{aligned}
&x' = \frac{1}{6}K_3 L\theta^3(l_b+l_d)^3\delta^3-\frac{1}{2}K_3 L\theta^2(l_b+l_d)^2\delta^2 \Delta x\\
&y' = -\frac{1}{2}K_3 L\theta^2(l_b+l_d)^2\delta^2 \Delta y
\end{aligned}
\end{equation}

The additional sextupole kick will remain imprinted on the transverse phase space causing additional emittance growth.  Provided the octupole kicks are cancelled, the emittance growth at the exit of BC2 from a misaligned octupole is approximately given by:
\begin{equation}
\begin{aligned}
&\frac{\Delta \epsilon_x}{\epsilon_x} \approx \frac{3}{8}\frac{\beta_{x}}{\epsilon_{x}}\bigg( K_{3}L{\theta}^2(l_{b}+l_{d})^2{\sigma_{\delta}}^2 \Delta x \bigg)^2\\
&\frac{\Delta \epsilon_y}{\epsilon_y} \approx \frac{3}{8}\frac{\beta_{y}}{\epsilon_{y}}\bigg( K_{3}L{\theta}^2(l_{b}+l_{d})^2{\sigma_{\delta}}^2 \Delta y \bigg)^2\\
\end{aligned}
\end{equation}
Here $l_b$, $l_d$, $\theta$, $K_3$ and $L$ refer to the BC1 or BC2 chicane and octupole parameters, $\beta_{x,y}$ and $\epsilon_{x,y}$ are the beta function and geometric emittance at the octupole, and $\sigma_{\delta}$ is the RMS energy spread at the chicane entrance.  This expression gives 10\% x emittance growth for $\pm$100 $\mu$m offset of the BC1 and BC2 octupoles and 10\% y emittance growth  for $\pm$50 $\mu$m offset of the BC1 octupole and $\pm$100 $\mu$m of the BC2 octupole. 

These alignment tolerances are relaxed when we consider emittance growth in the core of the beam including the additional emittance growth observed in {\footnotesize ELEGANT} simulations.  Figure 8, shows the emittance growth varying the x and y offset of the BC1 and BC2 octupoles in {\footnotesize ELEGANT}.  From this we see 10\% y-emittance growth for $\pm$150 $\mu$m offset of the BC1 octupole and $\pm$200 $\mu$m for the BC2 octupole and 10\% x-emittance growth for $\pm$200 $\mu$m offset of the BC1 and BC2 octupoles.  We also observe that the remnant emittance growth caused by 2nd order chromatic focusing is reduced by the sextupole kick from x offset of the octupoles.  Furthermore, the x-emittance growth due to BC1 sextupole kick is corrected by an opposite x offset of the BC2 octupole.  This effect can possibly be utilized to reduce the final emittance growth.

\section{conclusion}
The double octupole scheme demonstrates effective suppression of current horns by adjusting the higher order compression of a linac with multi-stage bunch compression.  This allows for the generation of a flat current profile with an increase in the achievable peak current and preservation of the transverse beam quality.  This method can be incorporated in most existing high brightness linacs, and could be improved upon if considered in the initial design of such a facility.  In the provided LCLS-II example, for an octupole length, $L=0.2$ m, the octupole strengths quoted throughout correspond to modest pole tip fields of 1.4 kG and 0.33 kG with an aperture of 70 mm and 50 mm respectively, with both values larger than the beam clearance requirement of the corresponding bunch compressor.

\begin{acknowledgments}
The authors would like to thank Agostino Marinelli, Tessa Charles and Paul Emma for useful discussion. This work was
supported by U.S. Department of Energy Contract No. DE-AC02-76SF00515.
\end{acknowledgments}

\begin{figure}[H]
\centering
\includegraphics[scale=0.5]{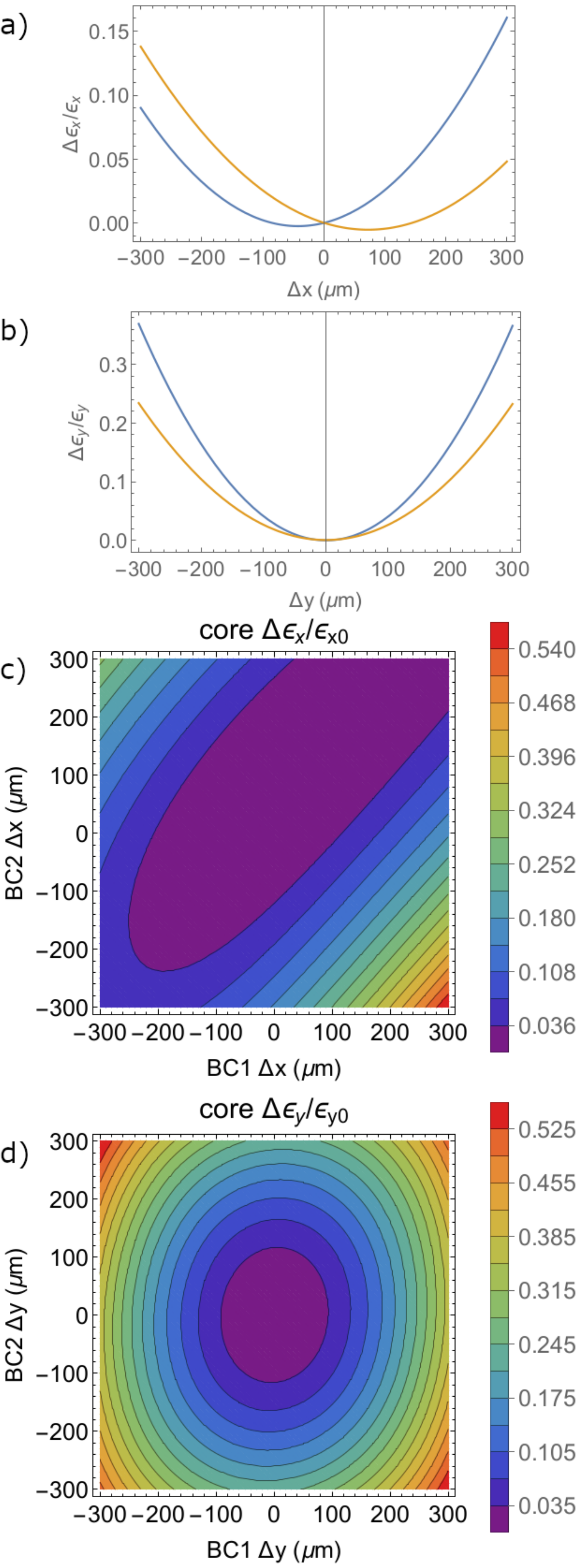}
\caption{a) The x normalized emittance at the BC2 exit of the core of the beam varying the BC1 octupole x offset (blue) and BC2 octupole x-offset (yellow) b) The y normalized emittance at the BC2 exit of the core of the beam varying the BC1 octupole y offset (blue) and BC2 octupole y-offset (yellow). c) The change in x normalized emittance at the BC2 exit of the core of the beam varying both BC1 and BC2 x-offset. d) The change in y normalized emittance at the BC2 exit of the core of the beam varying both BC1 and BC2 y-offset.}
\label{layout}
\end{figure}


\begin{thebibliography}{99}
\bibitem{FEL2}
P. Neyman, W. B. Colson, S. C. Gottshalk, A. M. M. Todd,
J. Blau, and K. Cohn, Free electron lasers in 2017,
in {\it Proceedings of the 38th International Free Electron
Laser Conference (FEL’17), Santa Fe, NM}, 2017 (JACoW,
Geneva, 2018), pp. 204–209.

\bibitem{RF}
H. Wiedemann, Particle Accelerator Physics (SpringerVerlag, Berlin), 3rd ed. (2007)

\bibitem{LSC}
M. Venturini, Phys. Rev. ST Accel. Beams {\bf 11}, 034401, (2008)

\bibitem{CSR1}
J. B. Murphy, {\it An Introduction to Coherent Synchrotron Radiation in Storage Rings}, ICFA Beam Dynamics Newsletter No. 35, p. 20-27 (2004)

\bibitem{CSR3}
K. L. F. Bane et al., Phys. Rev. ST Accel. Beams {\bf 12} 030704 (2009)

\bibitem{WAKE1}
K. F. Bane, Report No. SLAC-PUB-11829, (2006)

\bibitem{WAKE2}
A. Novokhatski and A. Mosnier, Nucl. Instrum. Methods Phys. Res., Sect. A {\bf 763}, 202, (2014)

\bibitem{HARM}
P. Emma, LCLS technical note, Report No. SLAC-TN-05-004, (2001)

\bibitem{OPT1}
Y. Sun, P. Emma, T. Raubenheimer, and J. Wu, Phys. Rev. ST Accel. Beams {\bf 17}, 110703 (2014)

\bibitem{OPT2}
S. Di Mitri, M. Cornacchia, and S. Spampinati, Phys. Rev. Lett. {\bf 110}, 014801 (2013)

\bibitem{OPT3}
C. Mitchell, J. Qiang, and P. Emma, Phys. Rev. ST Accel. Beams, {\bf 16}, 060703 (2013)

\bibitem{OPT4}
G. Penco et al., Phys. Rev. Lett. {\bf 112}, 044801 (2014)

\bibitem{OPT5}
G. Penco et al., Phys. Rev. Lett. {\bf 119}, 184802 (2017)

\bibitem{OPT6}
H. X. Deng et al., Phys. Rev. Lett. {\bf 113}, 254802 (2014)

\bibitem{OPT7}
Y. Ding et al., Report No. LCLS-II-TN-18-02 (2018)

\bibitem{LCLS}
J. Arther et al., Linac Coherent Light Source (LCLS) Conceptual Design Report No. SLAC-R-593, 2002

\bibitem{SWISSFEL}
B. Beutner, {\it Proc. of the 4th International Particle Accelerator Conf., IPAC 2013, Shanghai, China}, WEPFI057, p. 281

\bibitem{DIMITRI}
S. Di Mitri and M. Cornacchia, Phys. Rep. {\bf 539}, 1 (2014)

\bibitem{LCLSII}
T. Raubenheimer et al., {\it LCLS-II final design report. Technical report}, SLAC Technical Report. No. LCLSII1.1-DR-0251-R0, (2015)

\bibitem{DING}
Y. Ding et al., Phys. Rev. Accel. Beams {\bf 19}, 100703 (2016)

\bibitem{TESSA1}
T. K. Charles et al., Phys. Rev. Accel. Beams, {\bf 20}, 030705 (2017)

\bibitem{TESSA2}
T. K. Charles, D. M. Paganin and R. T. Dowd, Phs. Rev. Accel. Beams, {\bf 19} 104402 (2016)

\bibitem{TESSA3}
Y. Ding, K. Bane and Y. Nosochkov, {\it Proc. of 39th Free Electron Laser Conf., FEL 2019, Hamburg, Germany}, THP035

\bibitem{ELEGANT}
M. Borland, ANL Advanced Photon Source Report No.  LS-287, (2000)

\bibitem{CHAO}
 A. W. Chao and M. Tigner, {\it Handbook of Accelerator Physics and Engineering} (World Scientific, Singapore, 1999)

\bibitem{IMPACT1}
J. Qiang, R. D. Ryne, S. Habib, and V. Decyk, J. Comput. Phys. {\bf 163}, 434 (2000)

\bibitem{IMPACT2}
J. Qiang, S. Lidia, R. D. Ryne, and C. Limborg-Deprey, Phys. Rev. ST Accel. Beams {\bf 9}, 044204 (2006)

\bibitem{YURI}
Y. Nosochkov, Report No. LCLS-II-TN-20-04, (2020)

\bibitem{NICK}
N. Sudar, K. Bane, Y. Ding, Y. Nosochkov and Z. Zhang, Report No. LCLS-II-TN-20-05 (2020)

\end{thebibliography}
\end{document}